\documentclass[fleqn,12pt,oneside]{article}
\usepackage{espcrc1}

\usepackage{graphicx}
\usepackage[figuresright]{rotating}

\def\PRL{Phys. Rev. Lett.}
\def\PLB{Phys. Lett. B}

\def\etal{\emph{et al.}}

\title{Jets and Dijets in Au+Au and p+p Collisions at RHIC}

\author{David Hardtke\address{Nuclear Science Division\\ 
        Lawrence Berkeley National Laboratory\\
        Berkeley, CA  94720} 
        for the STAR Collaboration\footnote{For the full author list and acknowledgements, see Appendix ``Collaborations'' of this volume.}}%

\begin{document}

\maketitle

\begin{abstract}

Recent data from RHIC suggest novel nuclear effects in the production of
high $p_T$ hadrons.  We present results from the STAR detector 
on high $p_T$ angular correlations in Au+Au and p+p collisions 
at $\sqrt{s}=200$ GeV/c.  
These two-particle angular correlation measurements verify the presence of
a partonic hard scattering and fragmentation component at high $p_T$ in 
both central and peripheral Au+Au collisions.  When triggering on a leading
hadron with $p_T>$4 GeV, we observe a quantitative agreement between the 
jet cone properties in p+p and all centralities of Au+Au collisions.  
This quantitative agreement indicates that nearly all hadrons with 
$p_T>$4 GeV/c come from jet fragmentation and that jet fragmentation 
properties are not substantially modified in Au+Au collisions.  
STAR has also measured the
strength of back-to-back high $p_T$ charged hadron correlations, 
and observes a small suppression of the back-to-back correlation strength in 
peripheral collisions, and a nearly complete disappearance of back-to-back 
correlations in central Au+Au events.  These phenomena, toghether with the 
observed strong suppression of inclusive yields and large value of 
elliptic flow at high $p_T$, are consistent
with a model where high $p_T$ hadrons come from partons created near the 
surface of the collision region, and where partons that originate or propagate
towards the center of the collision region are substantially slowed or 
completely 
absorbed.

\end{abstract}

\section{Introduction}

At high energy density, nuclear matter may undergo a phase 
transition to a deconfined state consisting of free quarks and gluons.  This
Quark-Gluon Plasma may be experimentally accessible using the collisions
of heavy nuclei at high energies.  In order to investigate this new state of 
matter, processes whose rates are calculable in the absence of nuclear effects
are particularly useful.  The production of large transverse momentum jets, in
the limit of no nuclear effects, should scale with the number of binary 
nucleon-nucleon collisions with a rate calculable from perturbative QCD.  
A deviation from this binary scaling expectation for jet production
or a modification of jet properties would indicate the presence of
initial and/or final state nuclear effects.  The final state nuclear
effects are of particular interest since they may be used to verify the
presence of a Quark-Gluon plasma and may probe the properties of the plasma.

A large momentum parton traversing a dense gluonic system may 
lose energy due to collisions or 
induced gluon radiation \cite{Gyulassy,Wang}.  
The rate of energy
loss ($dE/dx$) is expected to scale with the density of the gluonic system.
Both PHENIX and STAR have measured the single inclusive hadron yields at high
$p_T$ and found a suppression compared to the binary nucleon-nucleon scaling 
expectation \cite{PHENIX_highpt,STAR_highpt}.  While these measurements are consistent with energy loss in
a dense gluonic system, the particle production mechanism for high $p_T$ 
hadrons in central Au+Au collisions may be different than in elementary 
collisions.  Indeed, the ratio of proton to pion at 
large $p_T$ ($\approx 2-4$ GeV/c) has been measured by the 
Phenix collaboration to be near unity in central Au+Au collisions, 
compared to a ratio of $\approx 0.2$ measured in p+p collisions \cite{Phenix_ppi}.  This has
been interpreted by some as an indication of a large hydrodynamical type 
contribution to the charged hadron yield in Au+Au collisions, even at 
rather large $p_T$.  The present work addresses the particle production
mechanism at high $p_T$ using two-particle azimuthal correlations.  We show
that hard scattering and fragmentation is the dominant particle production
mechanism for $p_T>$4 GeV/c.  In addition, we show that back-to-back 
dihadrons, a signature of dijets, are suppressed in the most central 
Au+Au collisions.  

\section{Experiment and Method}

Partons fragment into jets of hadrons around the direction of parton 
propagation.  Due to the large multiplicities in central Au+Au collisions
at RHIC, full jet reconstruction is difficult. To identify
jets on a statistical basis, STAR utilizes two-hadron azimuthal correlations
at large transverse momentum. The STAR experiment has been described in
detail elsewhere \cite{STARNIM}, so only details particularly relevant to this analysis are given.  The main tracking detector is a large acceptance
Time Projection Chamber with full azimuthal coverage and large pseudo-rapidity
acceptance ($|\eta| < 1.5$).  The TPC resides in a 0.5T solenoidal magnet.
The trajectories of charged tracks are measured in the TPC, 
and the interaction 
vertex is reconstructed on an event-by-event basis.  The tracks in the TPC
are projected to the primary vertex, and those passing within 1 cm of the 
reconstructed primary vertex are used in the analysis.  The detector
has excellent momentum resolution (the Gaussian width of
the track curvature $k \propto 1/p_T$ is $\delta k/k = 0.005(p_T/(GeV/c))+0.0076$), and many sources of 
secondary tracks (conversions, weak-decays, etc.) are rejected
due to the vertex constraint.   

For this analysis, we utilize $\approx 10$ million minimum bias p+p and
$\approx 1.7$ million minimum bias Au+Au events.  To augment the central
data sample, an additional 1.5 million events triggered on 
the $\approx$10\% most
central collisions are used.  For the analysis, we identify events with at 
least one large transverse momentum track, which we call the trigger particle.
The analysis is done for trigger particle thresholds $3<p_T^{trig}<4$, 
$4<p_T^{trig}<6$, and $6<p_T^{trig}<8$ GeV/c.  An azimuthal distribution
is constructed for all other charged 
tracks in these events with 2 GeV/c $<p_T<p_T^{trig}$,
\begin{equation} C_2(\Delta \phi) =
\frac{1}{N_{trigger}}\frac{1}{\epsilon} \int d\Delta \eta N(\Delta \phi, \Delta 
\eta),
\end{equation} 
where $N_{trigger}$ is the observed number of tracks
satisfying the trigger requirement, and $N(\Delta \phi, \Delta \eta)$
is the number of trigger-associated particle pairs 
as a function of their relative azimuth
($\Delta \phi$) and pseudo-rapidity ($\Delta \eta$), and $\epsilon$ is the 
efficiency for finding the associated track.  The analysis is restricted to 
the range $|\eta|<0.7$ ($|\Delta \eta|<$1.4).   

\section{Fragmentation Contribution to High $p_T$ Particle Production}

\begin{figure}[htb]
\begin{center}
\includegraphics[height=8cm]{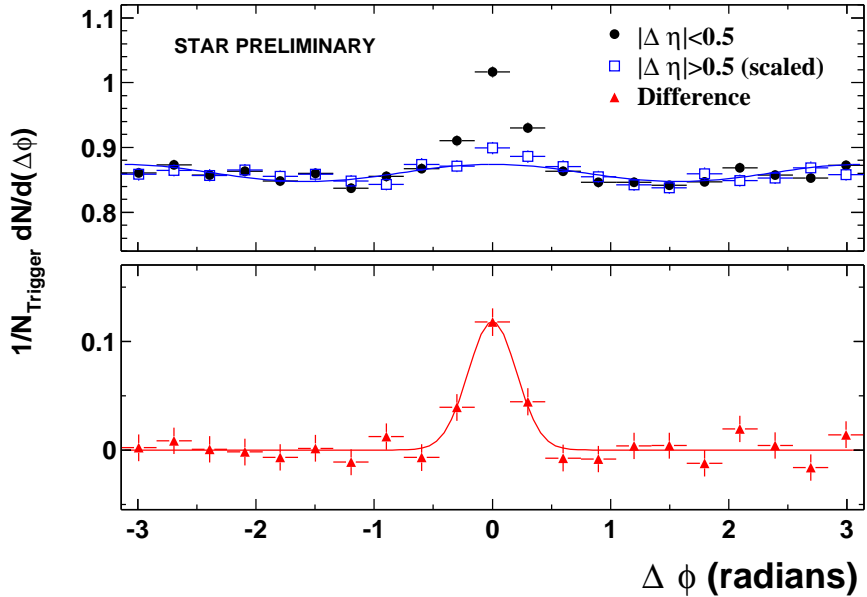}
\end{center}
\caption{Charged hadron azimuthal distributions for 
central Au+Au collisions at$\sqrt{s_{NN}} = 200$ GeV/c. 
The trigger particle threshold is $4<p_T^{trig}<6$ GeV/c and
the associated particle threshold is $2$ GeV/c $<p_T<p_T^{trig}$.
The circles are for $|\Delta \eta|<0.5$, the squares are for 
$0.5<|\Delta \eta|<1.4$, and the triangles represent the difference.}
\label{hist_backsub}
\end{figure}

In order to verify the presence of a hard scattering and fragmentation 
component at high $p_T$, small-angle azimuthal correlations 
are used \cite{STARv2}.  In addition to jets, however, there are other sources
of azimuthal correlations.  In both p+p and Au+Au collisions, there are 
azimuthal correlations due to momentum conservation, dijets, and resonance
decays.  Unique to Au+Au collisions are azimuthal correlations due to elliptic
flow.  The azimuthal correlations due to momentum conservation, dijets, and
elliptic flow should have only a weak pseudo-rapidity dependence within the 
acceptance window used for this analysis ($|\eta|<0.7$).  To 
disentangle the sources of azimuthal correlations, the azimuthal distributions
can be measured as a function of relative pseudo-rapidity.  In the top panel of Figure \ref{hist_backsub}, the azimuthal distributions for central Au+Au
collisions are shown for $|\Delta \eta|<0.5$ and $0.5<|\Delta \eta|<1.4$.  
The small relative pseudo-rapidity azimuthal distribution is absolutely
normalized, while the large relative pseudo-rapidity distribution is scaled
to match in the region where little correlated jet or dijet contribution 
is expected ($0.75<|\Delta \phi|<2.24$ radians).  There is a clear
excess near $\Delta \phi = 0$ seen at small relative pseudo-rapidity.  The
bottom panel of Fig. \ref{hist_backsub} shows the difference in the two
distributions.  The difference shows the same excess near $\Delta \phi = 0$,
while showing a flat distribution at larger relative azimuth.  From this
we conclude that there are correlations localized in 
$\Delta \phi, \Delta \eta$.  These correlations are expected from jets
and/or resonances.

\begin{figure}[htb]
\begin{center}
\includegraphics[height=8cm]{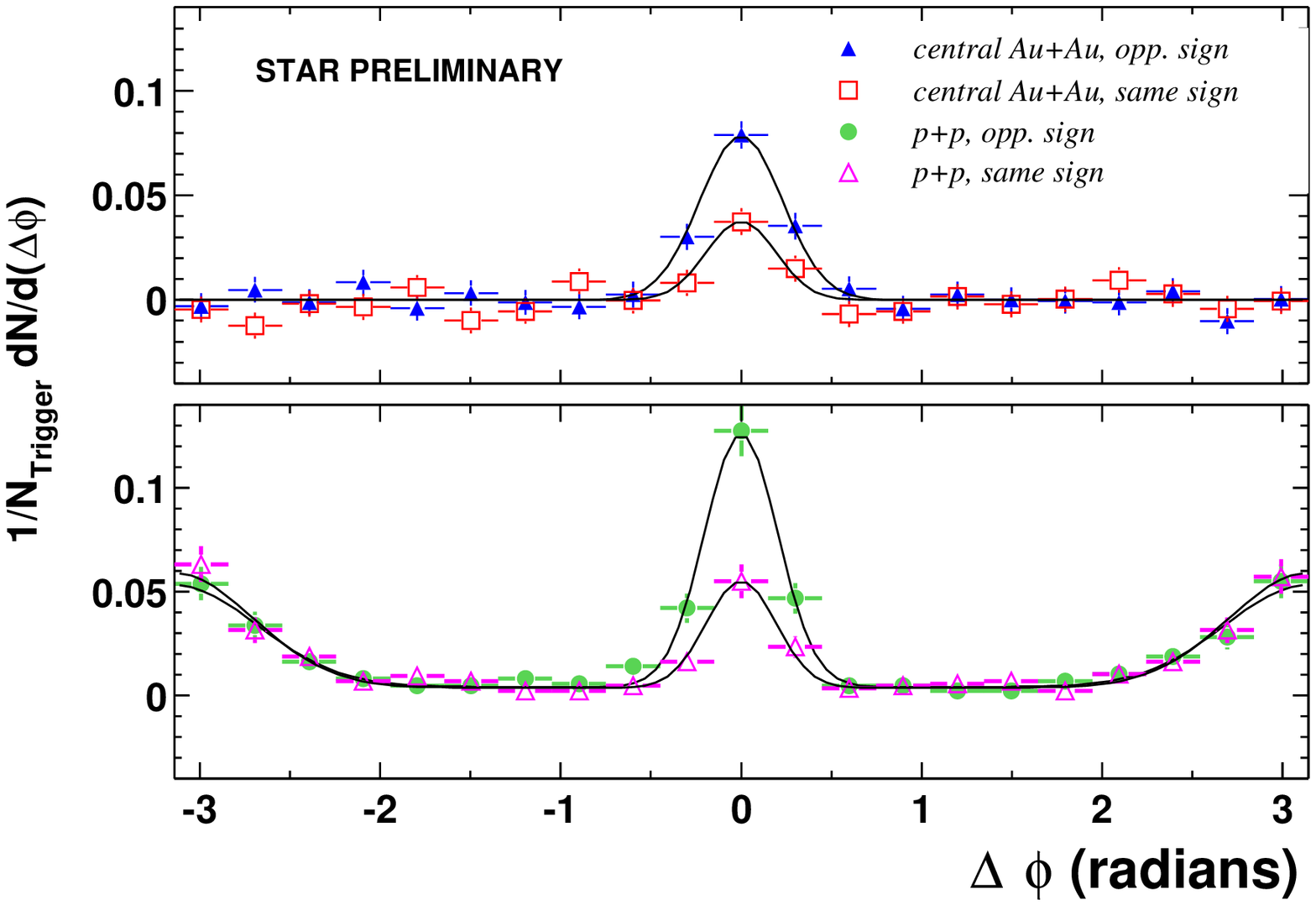}
\end{center}
\caption{Charged hadron azimuthal distributions for same sign and
opposite sign pairs. The top panel shows 
central Au+Au data ($|\Delta \eta|<0.5-0.5<|\Delta \eta|<1.4$(scaled)),
and the bottom panel shows p+p data ($0<|\Delta \eta|<1.4$).
The trigger particle threshold is $4<p_T^{trig}<6$ GeV/c and
the associated particle threshold is $2$ GeV/c $<p_T<p_T^{trig}$.}
\label{chargesign}
\end{figure}

In order to elucidate the nature (i.e. jets versus resonances) 
of the small angle correlations in 
Fig.\ref{hist_backsub}, we exploit a known property of jet fragmentation.  
Dynamical charge correlations exist within jet fragmentation \cite{Delphi}.
Leading and next-to-leading charged tracks tend to have opposite charge
signs.  Figure \ref{chargesign} shows the azimuthal distributions for minimum
bias p+p and central Au+Au data for opposite and same sign particle pairs.  
The p+p data is for $0<|\Delta \eta|<1.4$. The central Au+Au data
is for  $(|\Delta \eta|<0.5)-N*(0.5<|\Delta \eta|<1.4)$, thereby removing
azimuthal correlations due to dijets and elliptic flow.  For both the 
p+p data and central Au+Au data, the correlation strength near    
$\Delta \phi = 0$ is larger for opposite sign pairs than for same sign pairs.
Integrating the correlation excess, the ratio of opposite sign to same sign
peak areas is 2.7$\pm$0.6 for p+p data and 2.4$\pm$0.6 for central Au+Au data.
Phenomenological calculations using Pythia 
\cite{lund} (based on the Lund string 
fragmentation scheme) predict a ratio of 2.6$\pm$0.7. Large transverse
momentum particle production in p+p collisions (and as implemented in the 
Pythia model) is dominated by hard scattering and fragmentation, so the
agreement between central Au+Au, p+p, and Pythia calculations suggests
the same underlying particle production mechanism.  In addition, the finite
same sign azimuthal correlations disfavor the resonance decay hypothesis since
only the $\Delta$ resonance can decay into same sign particle pairs.     

\section{Dijets in Au+Au collisions}

\begin{figure}[htb]
\begin{center}
\includegraphics[height=11cm]{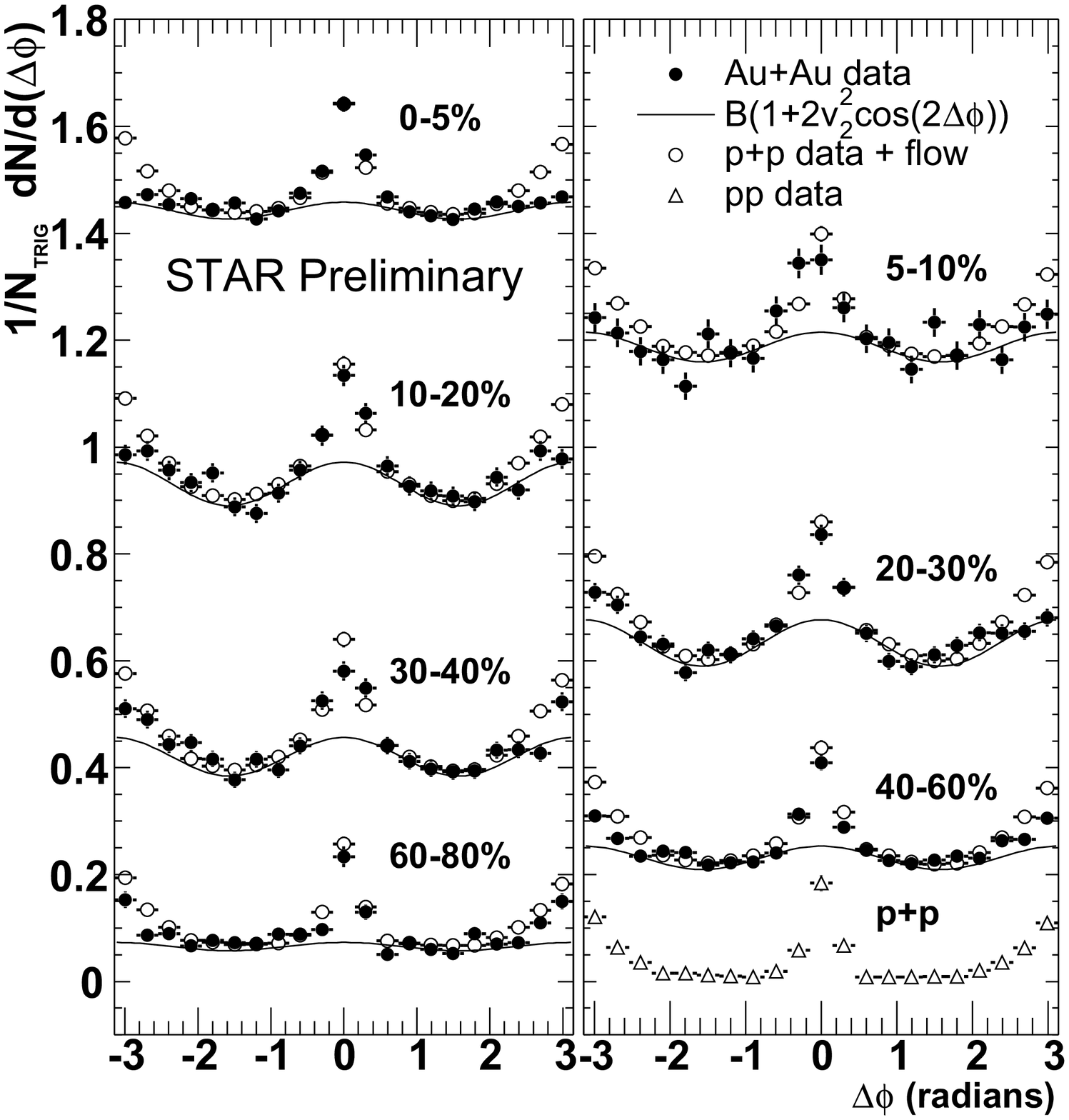}
\end{center}
\caption{Azimuthal distributions ($0<|\Delta \eta|<1.4$) for Au+Au 
collisions (solid circles) compared to
the expected azimuthal distributions from 
Equation \protect{\ref{c2eqn}}(open circles).  
The trigger particle threshold is $4<p_T^{trig}<6$ GeV/c and
the associated particle threshold is $2$ GeV/c $<p_T<p_T^{trig}$.
Also shown is the expected elliptic flow contribution for each
centrality (solid curve).}
\label{data}
\end{figure}

Figure \ref{chargesign} also shows clear evidence for back-to-back dihadrons
within the STAR acceptance for p+p collisions at $\sqrt{s}=200$ GeV/c.  The
presence of back-to-back dihadrons is an indication of dijets.  To compare
the dihadron production rates in Au+Au and p+p collisions, we 
treat high $p_T$ triggered Au+Au
events as a superposition of a high $p_T$ triggered p+p collision 
collision and a combinatorical elliptic flow background,
\begin{equation}
C_2^{\mathrm{AuAu}} = C_2^{\mathrm{pp}} + B(1+2v_2(p_T^t)v_2(p_T^a) \cos(2\Delta
 \phi)),
\label{c2eqn}
\end{equation}
where $p_T^t(p_T^a)$ is the trigger (associated) particle transverse momentum.
The elliptic flow parameters $v_2$ are measured independently using a reaction plane
method \cite{Filimonov}.  Since $v_2$ is approximately constant for $p_T>$2 GeV/c, a constant value is used.  The parameter $B$ is determined by fitting in the region
$0.75<|\Delta \phi|<2.24$.
In Figure \ref{data}, we compare the azimuthal 
distributions in Au+Au at various centralities to the expectation from 
Eqn. \ref{c2eqn}.  Qualitatively, the Au+Au azimuthal distributions at all centralities are described quite well near $\Delta \phi = 0$ as the superposition of elliptic flow and p+p azimuthal distributions.  In contrast, the back-to-back
azimuthal correlations predicted from Eqn. \ref{c2eqn} are always larger
than the back-to-back correlations seen in the Au+Au data.  In the most
peripheral Au+Au collisions, the suppression of 
back-to-back dihadrons is rather small, while
for the most central collisions there is little indication of any 
back-to-back dihadrons beyond those expected from elliptic flow. 

\begin{figure}[htb]
\begin{center}
\includegraphics[height=8cm]{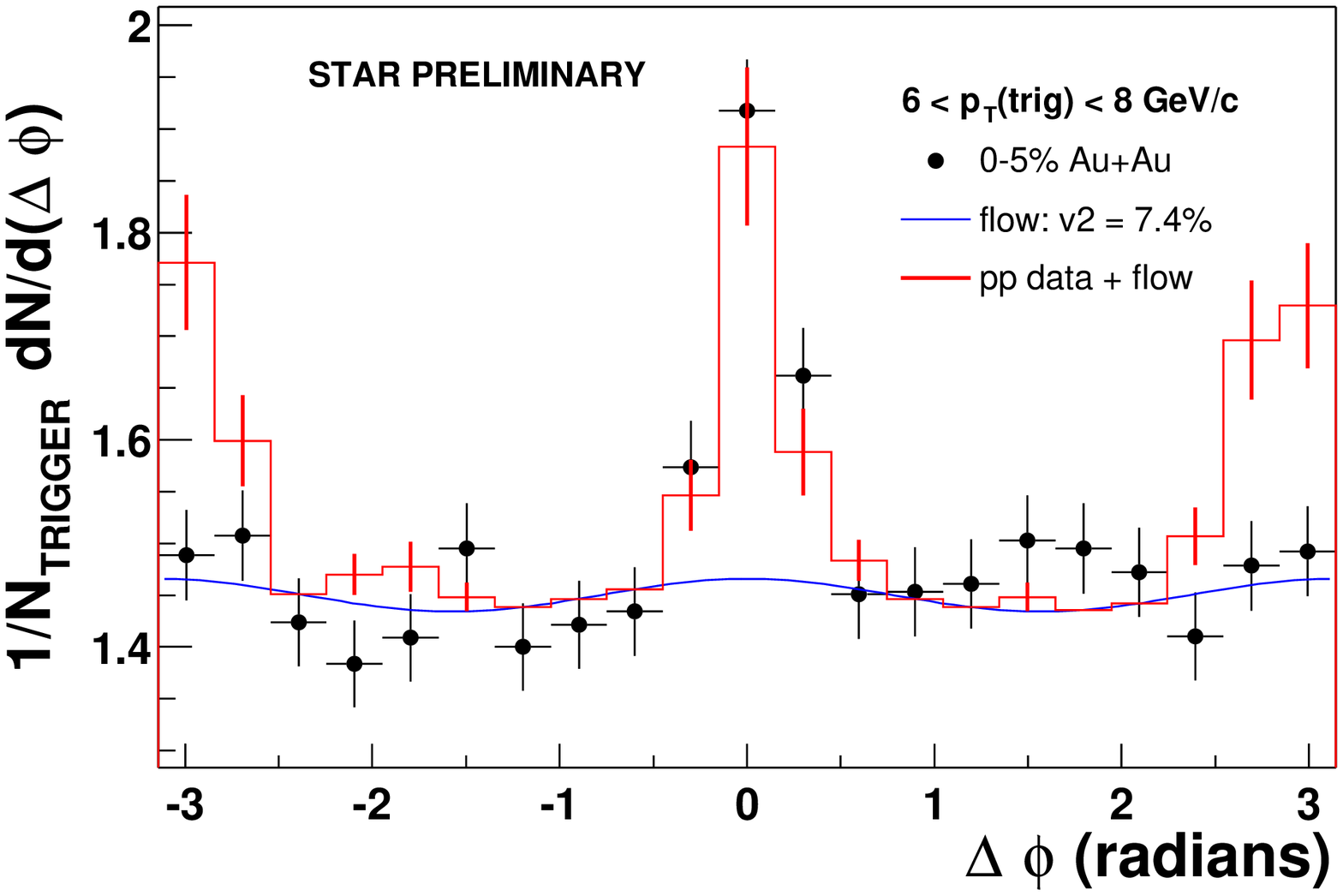}
\end{center}
\caption{Azimuthal distributions ($0<|\Delta \eta|<1.4$) for central
Au+Au 
collisions (solid circles) compared to
the expected azimuthal distributions from 
Equation \protect{\ref{c2eqn}}(histogram).  
The trigger particle threshold is $6<p_T^{trig}<8$ GeV/c and
the associated particle threshold is $2$ GeV/c $<p_T<p_T^{trig}$.
Also shown is the expected elliptic flow contribution (solid curve).}
\label{data6GeV}
\end{figure}

As the trigger particle threshold is increased, it might be expected that
back-to-back dihadrons reappear.  In Figure \ref{data6GeV}, the azimuthal
distribution expected from Eqn. \ref{c2eqn} is compared to that measured
in central Au+Au events for $6<p_T^{trig}<8$ GeV/c.  Once again, the small
angle azimuthal distribution measured in Au+Au is well described by Equation
\ref{c2eqn}, while there is an absence of back-to-back correlated 
dihadrons in the central Au+Au data.

\begin{figure}[htb]
\begin{center}
\includegraphics[height=10cm]{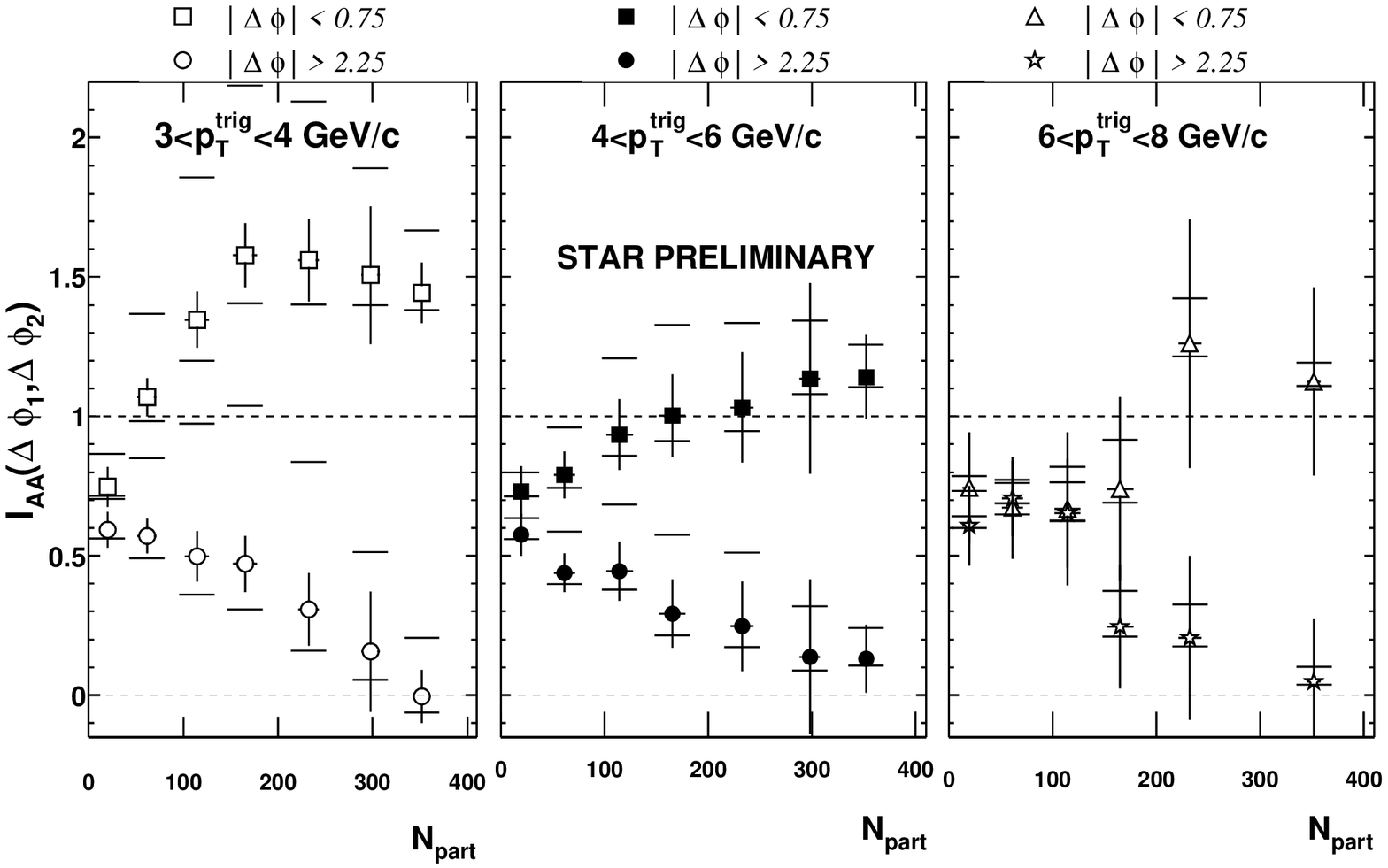}
\end{center}
\caption{$I_{AA}$ versus number of participants for small-angle ($|\Delta \phi|<0.75$) and back-to-back ($|\Delta \phi|>2.24$) azimuthal regions.}
\label{IAA}
\end{figure}

To quantify deviations from Eqn. \ref{c2eqn}, we form a ratio of the Au+Au
correlation excess beyond that expected from elliptic flow and the p+p correlation excess,
\begin{equation}
I_{AA}(\Delta \phi_1,\Delta \phi_2) = 
\frac{\int_{\Delta \phi_1}^{\Delta \phi_2} d(\Delta \phi) [C_2^{\mathrm{AuAu}}
- B(1+2v_2^t v_2^a \cos(2 \Delta \phi))]}{\int_{\Delta \phi_1}^{\Delta \phi_2} d
(\Delta \phi) C_2^{\mathrm{pp}}}.
\label{ratioeqn}
\end{equation}
This ratio is plotted as function of the number of participating nucleons
for the small-angle and back-to-back azimuthal regions in Figure \ref{IAA}.
The left panel shows the ratios for $3<p_T^{trig}<4$ GeV/c, the middle
panel shows the ratios for $4<p_T^{trig}<6$ GeV/c, and the right panel shows
the ratios for $6<p_T^{trig}<8$ GeV/c.  The horizontal bars show the 
systematic error on the ratio due the +5/-20\% systematic uncertainty
on the $v_2$ measurement from the reaction plane method \cite{STARv2}.  
For all three trigger particle thresholds, $I_{AA}(0,0.75)$ is smaller than unity
for the most peripheral collisions, and increases with increasing number
of participants.
The dependence on centrality is strongest for the lowest trigger particle 
$p_T$ threshold.  In contrast, $I_{AA}(2.24,\pi)$ decreases with increasing
$N_{part}$, and approaches zero for the most central collisions for all
three trigger particle thresholds. Note that with increasing trigger particle 
threshold, the jet and dijet correlation signals increase relative to the 
elliptic flow contribution, leading to smaller systematic errors on the 
$I_{AA}$ determination.   

It has been shown in fixed target experiments at Fermilab that the shape
of the back-to-back dihadron azimuthal distribution is sensitive to the 
intrinsic parton $k_T$ \cite{Apanasevich}.  In proton-nucleus and 
nucleus-nucleus collisions, additional initial state transverse momentum is
generated due to multiple nucleon-nucleon interactions.  To investigate
whether this nuclear $k_T$ can cause the observed deficit of back-to-back
azimuthal correlations in Au+Au collision, the $k_T$ parameter in Pythia
was varied from the nominal value of $\sigma = 1$ GeV/c to a
value of $\sigma = 4$ GeV/c.  This was found to have only a small effect
on $I_{AA}(2.24,\pi)$, resulting in less than a 20\% reduction in the predicted
value.

\section{Discussion and Conclusions}

These data on azimuthal correlations and other
data on high $p_T$ particle production at RHIC can be explained
with a simple model.  The single particle inclusive yields for $p_T> 2$ GeV/c
are suppressed in central Au+Au collisions 
compared to the binary nucleon-nucleon scaling expectation.
In addition, the elliptic flow has been shown to saturate above $p_T \approx
2-3$ GeV/c.  The measured small angle azimuthal correlations suggest
that the properties of the jets we do observe via their leading particles
are substantially unmodified by the nuclear medium.  These jets most likely come
from partons produced near the surface of the collision region.  The suppression
of single particle inclusive yields and the disappearance of back-to-back
dihadron pairs in the most central collisions suggest that partons produced
in the center of the collision region, or their hadronic fragments,
 experience large energy loss or complete
absorption.  This surface emission scenario was predicted to
be a consequence of energy loss \cite{Bjorken}.

In summary, STAR has measured azimuthal correlations among high $p_T$ charged
hadrons.  The small angle correlations provide direct evidence for the
dominance of hard scattering and fragmentation processes in high $p_T$
particle production in Au+Au collisions at RHIC.  The observed relative
charge sign dependence is consistent with the expectation of parton 
fragmentation.  A simple model treating the azimuthal correlations as
a superposition of one hard scattering and elliptic flow is used
to compare the rates of back-to-back dihadron production in p+p and all 
centralities of Au+Au collision. The suppression
of back-to-back correlated dihadron pairs 
varies with centrality and is largest for the most central collisions.
These data suggest that high $p_T$ hadrons come from 
hard-scattering processes that occur near the edge of the collision region,
and that the medium produced in central Au+Au collisions at RHIC is 
opaque to the propagation of fast partons or their fragmentation products.

\end{document}